\documentclass[12pt]{article}
\usepackage{epsf, cite, amssymb}
\usepackage{epsfig}
\makeatletter
\renewcommand\section{\@startsection {section}{1}{\z@}%
                               {-3.5ex \@plus -1ex \@minus -.2ex}
                               {2.3ex \@plus.2ex}%
                               {\normalfont\large\bfseries}}
\renewcommand\subsection{\@startsection{subsection}{2}{\z@}%
                                 {-3.25ex\@plus -1ex \@minus -.2ex}%
                                 {1.5ex \@plus .2ex}%
                                 {\normalfont\bfseries}}
\makeatother

\def\IZ{\relax\ifmmode\mathchoice
{\hbox{\cmss Z\kern-.4em Z}}{\hbox{\cmss Z\kern-.4em Z}}
{\lower.9pt\hbox{\cmsss Z\kern-.4em Z}} {\lower1.2pt\hbox{\cmsss
Z\kern-.4em Z}}\else{\cmss Z\kern-.4em Z}\fi}
\def\IR{\relax{\rm I\kern-.18em R}}

\def\one{{\hbox{ 1\kern-.8mm l}}}

\newlength{\bredde}
\def\slash#1{\settowidth{\bredde}{$#1$}\ifmmode\,\raisebox{.15ex}{/}
\hspace*{-\bredde} #1\else$\,\raisebox{.15ex}{/}\hspace*{-\bredde}
#1$\fi}

\newsavebox{\zzzbar}
\sbox{\zzzbar}
{\setlength{\unitlength}{0.9em}
\begin{picture}(0.6,0.7)
\thinlines
\put(0,0){\line(1,0){0.6}}
\put(0,0.75){\line(1,0){0.575}}
\multiput(0,0)(0.0125,0.025){30}{\rule{0.3pt}{0.3pt}}
\multiput(0.2,0)(0.0125,0.025){30}{\rule{0.3pt}{0.3pt}}
\put(0,0.75){\line(0,-1){0.15}}
\put(0.015,0.75){\line(0,-1){0.1}}
\put(0.03,0.75){\line(0,-1){0.075}}
\put(0.045,0.75){\line(0,-1){0.05}}
\put(0.05,0.75){\line(0,-1){0.025}}
\put(0.6,0){\line(0,1){0.15}}
\put(0.585,0){\line(0,1){0.1}}
\put(0.57,0){\line(0,1){0.075}}
\put(0.555,0){\line(0,1){0.05}}
\put(0.55,0){\line(0,1){0.025}}
\end{picture}}

\newcommand{\ena}{\end{eqnarray}}
\newcommand{\beqa}{\begin{eqnarray}}
\newcommand{\eeqa}{\end{eqnarray}}
\newcommand{\bea}{\begin{eqnarray}}
\newcommand{\eea}{\end{eqnarray}}

\newcommand{\eq}[1]{(\ref{#1})}

\newcommand{\be}{\begin{equation}}
\newcommand{\ee}{\end{equation}}

\newcommand{\Tr}{{\rm Tr}}
\usepackage{graphicx}

\def\be{\begin{equation}}
\def\ee{\end{equation}}
\def\beq{\begin{eqnarray}}
\def\eeq{\end{eqnarray}}

\def\e{\epsilon}

\def\({\left (}
\def\){\right )}
\def\[{\left [}
\def\[{\right ]}

\def\ba{\begin{eqnarray}}
\def\ea{\end{eqnarray}}

\input amssym.def
\input amssym.tex
\begin{document}
\begin{titlepage}
\begin{flushright}
arXiv:0907.0889
\end{flushright}
\vfill
\begin{center}
{\LARGE\bf D-Brane Potentials from Multi-Trace\\
\vskip 3mm Deformations in AdS/CFT} \\
\vskip 10mm

{\large Alice Bernamonti%
\footnote{
Aspirant FWO
} and Ben Craps}
\vskip 7mm

Theoretische Natuurkunde, Vrije Universiteit Brussel, and \\
International Solvay Institutes \\
Pleinlaan 2, B-1050 Brussels, Belgium
\vskip 3mm 
\vskip 3mm 
{\small\noindent  {\tt Alice.Bernamonti@vub.ac.be, Ben.Craps@vub.ac.be}}
\end{center}
\vfill

\begin{center}
{\bf ABSTRACT}\vspace{3mm}
\end{center}
It is known that certain $AdS$ boundary conditions allow smooth initial data to evolve into a big crunch. To study this type of cosmological singularity, one can use the dual quantum field theory, where the non-standard boundary conditions are reflected by the presence of a multi-trace potential unbounded below. For specific $AdS_4$ and $AdS_5$ models, we provide a D-brane (or M-brane) interpretation of the unbounded potential. Using probe brane computations, we show that the $AdS$ boundary conditions of interest cause spherical branes to be pushed to the boundary of $AdS$ in finite time, and that the corresponding potential agrees with the multi-trace deformation of the dual field theory. Systems with expanding spherical D3-branes are related to big crunch supergravity solutions by a phenomenon similar to geometric transition.
\vfill
\end{titlepage}
\tableofcontents
\section{Introduction}

According to the AdS/CFT correspondence \cite{Maldacena:1997re}, type IIB string theory on global $AdS_5\times S^5$ with $N$ units of five-form flux, 
\be
\int_{S^5}\hat G_5=N,
\ee
is dual to ${\cal N}=4$ super-Yang-Mills (SYM) theory on $\mathbb{R}\times S^3$ with gauge group $SU(N)$. The string coupling $g_s$ is related to the Yang-Mills coupling $g_{YM}$ by $4\pi g_s=g_{YM}^2$, while the radius of curvature $R_{AdS}$ of both $AdS_5$ and $S^5$ is given in terms of the string length $l_s$ by
\be
{R_{AdS}^4\over l_s^4}=g_{YM}^2N.
\ee

The Poincar\'e patch of $AdS_5\times S^5$ with $N$ units of five-form flux appears as the near-horizon limit of $N$ coincident D3-branes. It is dual to ${\cal N}=4$ SYM theory on $\mathbb{R}^{1,3}$. This is an example of geometric transition \cite{Vafa:2000wi, Lin:2004nb}, where a spacetime with a number of branes (measured by the flux through a cycle surrounding the branes) is dual to a spacetime without branes but where the cycle has become non-contractible (the branes having been replaced by flux through the topologically non-trivial cycle). Since the D3-branes are BPS, they can be separated and placed at arbitrary positions in the dimensions transverse to their worldvolumes -- such configurations correspond to the Coulomb branch of ${\cal N}=4$ SYM theory on $\mathbb{R}^{1,3}$. D3-branes in $AdS_5\times S^5$ act as domain walls, separating regions that differ by one unit of five-form flux (and thus have different radius of curvature). If one wanted to decrease the five-form flux by one unit, one could send one of the D3-branes to infinity; similarly, to increase $N$, one could send in parallel D3-branes from infinity. Near-horizon limits of certain Coulomb branch configurations of D3-branes have been shown to lead to specific deformations of $AdS_5\times S^5$ \cite{Freedman:1999gk}.  

${\cal N}=4$ SYM theory on $\mathbb{R}\times S^3$ does not have a Coulomb branch: it is lifted by the conformal coupling of the scalar fields to the curvature of the $S^3$, which effectively makes those scalars massive. This is consistent with the fact that global $AdS$ cannot be obtained as the near-horizon limit of parallel D3-branes. However, D3-branes can still be used to increase the five-form flux $N$: a spherical D3-brane sent in from infinity is a domain wall; it will dynamically shrink and annihilate, leaving behind an $AdS_5$ with one more unit of five-form flux. The fact that a spherical probe D3-brane shrinks is not simply a consequence of it having a tension, since, in the regime of large radius, the leading effect of the tension (proportional to the fourth power of the radius of the D3-brane) is cancelled, because of a BPS relation between D3-brane tension and charge, by a compensating effect due to the four-form potential. Rather, it is a subleading quadratic potential that causes the shrinking \cite{Seiberg:1999xz, Hubeny:2004cn}. The counterpart in the dual gauge theory is the quadratic potential due to the conformal coupling of the transverse scalar fields. 

With the standard supersymmetric boundary conditions, $AdS_5\times S^5$ is stable. It is known, though, that a modification of the standard boundary conditions can introduce instabilities, allowing $AdS_5\times S^5$ to tunnel into a cosmological spacetime with a big crunch singularity, i.e.\ a spacelike singularity reaching the boundary of $AdS$ in finite global time \cite{Hertog:2004rz}. In the dual gauge theory, the modification of the $AdS$ boundary conditions corresponds to adding an unstable double trace potential to ${\cal N}=4$ SYM, allowing operators to become infinite in finite time \cite{Aharony:2001pa, Witten:2001ua, Berkooz:2002ug, Craps:2007ch}. The aim of the present paper is to provide a D3-brane interpretation of this instability. We will show that with modified boundary conditions, a spherical probe D3-brane feels a negative quartic potential in addition to its positive quadratic potential. As a consequence, $AdS$ can nucleate spherical D3-branes that are subsequently stretched to infinite size in finite time.\footnote{In the context of the AdS/CFT correspondence, it was shown in \cite{Maldacena:1998uz} that a closely related effect develops for non-supersymmetric spherical branes violating the BPS bound. In our present situation, the branes classically saturate the BPS bound and the repulsive force is generated by quantum corrections sensitive to the boundary conditions.} Every spherical D3-brane that is nucleated and stretched to infinity leaves behind a spacetime with one less unit of five-form flux, which is thus more strongly curved than the original $AdS$. In a gravity approximation (which in reality breaks down when the radius of curvature becomes of order the string scale), the result is a big crunch singularity.

Recently, a similar duality has been proposed between $AdS_4$ compactifications of M-theory or type IIA string theory and ABJM theory, an ${\cal N}=6$ superconformal $U(N)\times U(N)$ Chern-Simons theory with opposite levels $k$ and $-k$, respectively \cite{Aharony:2008ug}. An unstable triple trace deformation of ABJM theory was studied in \cite{Craps:2009qc}; as in \cite{Hertog:2004rz}, the corresponding boundary condition in the bulk $AdS_4$ allows smooth initial data to evolve into a big crunch singularity. We will show that this instability corresponds to spherical M2-branes or D2-branes being stretched to infinity in finite time, due to a potential generated by the modified boundary conditions. 

In \cite{Craps:2007ch}, a double trace deformation of ${\cal N}=4$ SYM was used to obtain a dual field theory description of a big crunch singularity, and an attempt was made to use self-adjoint extensions to evolve the system beyond the big crunch. This model is now understood not to be under good computational control, though, and a new proposal in the context of ABJM theory will appear in \cite{CHTinprogress}. Our present work shows that finding a consistent self-adjoint extension amounts to specifying what happens when spherical D-branes are stretched to infinite radius in finite time.
While at present it is unclear what can be learned from this new perspective, we hope that it will turn out to be useful in addressing these hard questions. Simpler systems for which our results may be useful are the hairy black hole solutions of \cite{Hertog:2005hu}, which are dual to stable multi-trace deformations and should be related to spherical D-branes with finite radius.

This paper is organized as follows. In section~2, we study spherical probe D3-branes in global $AdS_5\times S^5$ with modified boundary conditions. In section~3, we do the same for spherical probe M2-branes or D2-branes in global $AdS_4$ compactifications. In appendix~A, we compute brane potentials in Poincar\'e coordinates, which supplement the computations in global coordinates in the main text.

\setcounter{equation}{0}
\section{D3-branes in $AdS_5\times S^5$}

In this section, we study spherical D3-branes in global $AdS_5\times S^5$. This theory allows a consistent truncation to five-dimensional gravity with a negative cosmological constant coupled to a single scalar field \cite{Gunaydin:1985cu}. This scalar field corresponds to quadrupole deformations of $S^5$ and saturates the Breitenlohner-Freedman bound \cite{Breitenlohner82}. We will compute the D-brane effective potential as a function of the boundary condition on this bulk scalar field. Specifically, we will focus on boundary conditions corresponding to a classically marginal double trace deformation of ${\cal N}=4$ SYM theory. 

In the Feynman diagrams of interest, the D-brane emits a virtual scalar particle, which then interacts with the boundary before being reabsorbed by the D-brane. To compute such diagrams, we need two main ingredients: the coupling of the D3-branes to the bulk scalar field, and the effect of the boundary condition on the bulk scalar field. The coupling can be obtained from the well-known D-brane action and from the consistent truncation ansatz expressing the ten-dimensional bulk fields in terms of the five-dimensional metric and scalar field. The effect of the boundary conditions can be computed in two different ways. On the one hand, a modification of the boundary condition corresponds to adding a boundary term to the bulk action, which gives rise to a (quadratic) vertex that should be included in Feynman diagrams. The advantage of this approach is that it extends to non-linear boundary conditions (such as the ones we will study in section~3). On the other hand, the (linear) boundary conditions we are focusing on in this section can be fully taken into account by using a modified propagator for the scalar field. This approach has the advantage that it effectively resums diagrams with arbitrarily many boundary vertices inserted.

From the point of view of the full string theory on $AdS_5\times S^5$, one might wonder whether it is sufficient to compute the effective potential in the truncated five-dimensional theory. In particular, spherical D3-branes can also emit and reabsorb many other fields, which are not described by the consistent truncation. The point is, however, that only the bulk scalar field of the consistent truncation is directly affected by the modified boundary conditions: contributions from emission and reabsorption of other fields are the same as in the standard supersymmetric theory. For our purposes, it is therefore justified to work within the framework of the consistent truncation.

In section~2.1, we review the basic setup, in particular the relation between modified boundary conditions in $AdS$ and unstable double trace deformations in SYM. In section~2.2, we use the consistent truncation ansatz to determine the couplings of a spherical D3-brane to the bulk scalar field of interest. In section~2.3, we compute the propagator of the bulk scalar field, for standard as well as modified boundary conditions. In section~2.4, we compute the D-brane effective potential in the two ways described above. In section~2.5, we use the Coulomb branch solutions of \cite{Freedman:1999gk} to provide additional evidence that the big crunch singularity in the supergravity solutions of \cite{Craps:2007ch} is due to branes being pushed to the conformal boundary of $AdS$.

\subsection{Setup}\label{setup5d}

Type IIB supergravity compactified on $S^5$ can be consistently truncated to five-dimensional gravity coupled to a single $SO(5)$ invariant scalar $\varphi$ \cite{Gunaydin:1985cu}. From the ten-dimensional point of view, $\varphi$ arises from an $SO(5)$ invariant quadrupole distortion of $S^5$. The bulk action reads
\be\label{5daction}
S =  \frac{V_{S^5}}{\kappa_{10}^2 } \int d^5 x\sqrt{-g} \left[ \frac R 2 - \frac 1 2 \partial_{\mu}\varphi \partial^{\mu}\varphi  + \frac{1}{4 R^2_{AdS}} \(15 e^{2 \gamma \varphi} +10 e^{-4 \gamma\varphi} - e^{-10 \gamma \varphi} \)\right]\,, 
\ee
where $\gamma = \sqrt{2/15}$, $2 \kappa_{10}^2  = (2\pi)^{7} \alpha^{\prime 4} g_s ^2$, and $V_{S^5} = \pi^3 R_{AdS}^5$ is the volume of the internal manifold. 
The potential reaches a negative local maximum when the scalar vanishes; this is the maximally supersymmetric $AdS_5$ state, corresponding to the unperturbed $S^5$ in the type IIB theory. At linear order around the $AdS$ solution, the scalar obeys the free wave equation with a mass that saturates the Breitenlohner-Freedman \cite{Breitenlohner82} bound%
\footnote{\label{MBF}In $d+1$ dimensions, $m^2_{BF} = - d^2/(4R_{AdS}^2)$.} $m^2 = -4/R_{AdS}^2$. With the usual supersymmetric boundary conditions, $AdS_5$ is stable.

In global coordinates, the $AdS_5$ metric takes the form
\be\label{globalAdS}
ds^2 = -\(1+ \frac{r^2}{R_{AdS}^2}\) dt^2 + \frac{dr^2}{1+ \frac{r^2}{R^2_{AdS}}} + r^2 d\Omega_3^2\,.
\ee
In all asymptotically $AdS$ solutions, the scalar $\varphi$ decays at large radius as 
\be \label{scalar}
\varphi(x,r) = \frac{\alpha(x) \ln \(r \mu\)}{r^2}+\frac{\beta(x)}{r^2}\,. 
\ee
Throughout this section, we denote the five-dimensional bulk coordinates by $(r,x)$, where $x$ collectively denotes the time and three angular coordinates. The arbitrary scale $\mu$, necessary to define the logarithm, will be chosen to be $\mu = 1/R_{AdS}$ for most of this section (we will comment on this after \eq{Veffpert}).  The standard boundary conditions on the scalar field would set $\alpha=0$. This choice preserves the full $AdS$ symmetry group and has empty $AdS$ as its stable ground state. However, in the mass range $m^2_{BF} \le  m^2 < m^2_{BF} + 1/R_{AdS}^2$, one can consider more general boundary conditions of the form \cite{Witten:2001ua}
\be
\alpha = -\frac{\delta W}{\delta \beta}\,,
\ee
where $W(\beta)$ is an arbitrary real smooth function. 

We will be interested in scalar field boundary conditions
\be \label{BCs}
\alpha (x) = f \beta(x),
\ee
where $f$ is an arbitrary constant (for $f>0$, smooth initial data can develop a big crunch singularity \cite{Craps:2007ch}). This boundary condition does not preserve supersymmetry and breaks the asymptotic $AdS$ symmetries to $\mathbb{R} \times SO(4)$ \cite{Henneaux:2006hk}.

To obtain the boundary condition \eq{BCs} from a variational principle, one adds boundary terms to the bulk action \eq{5daction}. To do this in a precise way, we provide an IR regulator in the bulk by restricting the radial coordinate to $0 \le  r \le \Lambda$. (Through the UV/IR correspondence, the location $\Lambda$ of the regularized boundary in the bulk will correspond to a UV cutoff in the dual field theory). 
The boundary condition $\alpha = f \beta$ is obtained from the boundary term in the variation of the action if we add to the scalar field action \eq{5daction} the term
\be
S_{bdy} = \frac{V_{S^5}}{\kappa_{10}^2 R_{AdS}}\int_{\partial} d^4 x \sqrt{g_{bdy}} \left[-1 + \frac{f}{2\(1+ f \ln \(\Lambda /R_{AdS}\)\)}\right]  \varphi^2. \label{bdry5d}
\ee 
The first term also appeared in \cite{Sever:2002fk} (see also \cite{Hartman:2006dy}).

The AdS/CFT correspondence states that type IIB string theory on global $AdS_5\times S^5$ with standard ($f=0$) boundary conditions is dual to ${\cal N}=4$ SYM theory on $\mathbb{R}\times S^3$. Because of the conformal coupling to the curvature of $S^3$, which we choose to be of unit radius, the six adjoint scalar fields $\Phi_j$ of ${\cal N}=4$ SYM effectively get masses $m^2=1$.%
\footnote{\label{confcoupling}For a $d$-dimensional boundary, the mass would be $m^2= \frac{d-2}{4(d-1)} R_{S^{d-1}}$, where $R_{S^{d-1}}$ is the Ricci scalar of $S^{d-1}$.}  
According to \cite{Witten:2001ua, Berkooz:2002ug, Sever:2002fk}, changing the boundary condition on $\varphi$ to \eq{BCs} with non-zero $f$ corresponds to adding a double trace potential to the SYM action, 
\be\label{Osquared}
S = S_0+\frac{ f }{ 2} \int {\cal O}^2,
\ee
where ${\cal O}$ is the dimension 2 chiral primary operator dual to $\varphi$,
\be \label{operator}
{\cal O}=c\,\Tr\left[\Phi_1^2 - {1\over5} \sum_{i=2}^6 \Phi_{i}^2\right]
\ee
with $c$ a normalization constant of order $1/N$ (in conventions such that the fields $\Phi_i$ have canonical kinetic terms). The coupling $f$ in \eq{Osquared} is classically marginal but in fact marginally relevant for $f>0$ \cite{Witten:2001ua}.

The duality between type IIB string theory on (the Poincar\'e patch of) $AdS_5\times S^5$ and ${\cal N}=4$ SYM theory (on $\mathbb{R}^4$) can be obtained by taking a decoupling limit of a system of coincident D3-branes \cite{Maldacena:1997re}. In this picture, the eigenvalues of $\Phi_j$ correspond to D-brane positions, and the double trace potential in \eq{Osquared} provides a quartic potential for these D-brane positions. For $f>0$, the potential is unbounded below and sufficiently strong to push eigenvalues to infinity in finite time. Global $AdS_5\times S^5$, which is the background of interest in \cite{Hertog:2004rz, Craps:2007ch}, does not straightforwardly appear as a near-horizon limit of D3-branes. One could still expect that there should be a similar D-brane interpretation of the unstable potential in \eq{Osquared}. It is natural to assume that spherical D3-branes will play a role in this, as in \cite{Seiberg:1999xz, Hubeny:2004cn}. The main purpose of the present paper is to make this picture precise by computing the effective potential felt by spherical probe D3-branes as a function of the boundary conditions.  

\subsection{Coupling of the bulk scalar field to spherical D3-branes} \label{coupling}

The fact that the action \eq{5daction} is a consistent truncation of type IIB supergravity compactified on $S^5$ means that with any solution of \eq{5daction} one can associate a solution of the full type IIB supergravity equations of motion. The lift to ten dimensions is explicitly given in \cite{Cvetic00}. Let  $F =  e^{\gamma \varphi}$ (with $\gamma$ defined after \eq{5daction}) and $\Delta = F \sin^2 \xi + F^{-5} \cos^2 \xi$, where $\xi$ is a coordinate in terms of which the metric of the unit sphere would read
\be\label{xi}
d\Omega^2_5 = d\xi^2 + \sin^2 \xi d\Omega^2_4\,,
\ee
with $0\le \xi \le \pi$.
The full ten-dimensional metric is 
\be
ds_{10}^2  = \Delta^{1/2} ds_5^2 + R_{AdS}^2 \left[ F^4 \Delta^{1/2} d\xi^2 + F^{-1}\Delta^{-1/2}\sin^2\xi d\Omega^2_4 \right]\,.\label{10dg}
\ee
The self-dual five-form $\hat{G}_5 = G_5 + \ast G_5$ is determined by
\be
G_5  = -  \frac{ U}{R_{AdS}} \epsilon_5 + 6  R_{AdS}  \sin \xi \cos \xi F^{-1} \ast dF \wedge d\xi\,, \label{G5}
\ee
where we have denoted
\be
U = -3 F^2 \sin^2 \xi + F^{-10} \cos^2 \xi - F^{-4} - 4 F^{-4} \cos^2\xi\,.
\ee
In (\ref{G5}), $\epsilon_5$ and $\ast$ are the five-dimensionals volume-form and dual. 

To compute the coupling of a spherical D3-brane to the scalar field $\varphi$, we consider a probe D3-brane in the ten-dimensional lifted solution. In our computation of the effective potential for the D3-brane radius, we will only need the source term for the bulk scalar,%
\footnote{To compare with the deformation \eq{Osquared} of SYM, we will also need the kinetic terms for the scalar fields on the D3-brane world-volume.} 
so we work in a linearized approximation of the coupled scalar-gravity system about the $AdS$ background. The action of the probe brane is
\be
S_{D3} = S_{DBI}+S_{WZ}=- \tau_3 \int d^4 x \sqrt{- \hat{G}} + \mu_3 \int \hat{ C}_4\,,\label{SD3}
\ee
where $\hat G$ is the determinant of the pull-back of the ten-dimensional metric to the D3-brane world-volume and $d \hat{C}_4 =\hat{G}_5$. The tension and charge are given by $\tau_3 =\mu_3 = \sqrt{\pi}/ \kappa_{10}$. In the static gauge, the Dirac-Born-Infeld action includes the terms
\be
S_{DBI} = -\tau_3 \int d^4 x \sqrt{-\hat{g}} \left[  1 - 5 \gamma \varphi \(\cos^2 \xi - \frac 1 5 \sin^2 \xi \) + \frac 1 2 g_{i j} \partial_a x^i \partial^a x^j \right] \,,\label{BI}
\ee
where $\hat{g}$ is the determinant of the pull-back of the five-dimensional metric $g_{\mu\nu}$ to the four-dimensional world-volume, the index $a$ labels the four coordinates along the D3-brane world-volume, the index $i$ runs over the six transverse dimensions, and $\gamma = \sqrt{2/15}$ was introduced in \eq{5daction}. We rewrite the Wess-Zumino action as an integral over the five-dimensional volume enclosed by the D$3$-brane
\be
S_{WZ} = \mu_3 \int_{V_5} \hat{G}_5 = \frac{\mu_3 }{R_{AdS}} \int d^5 x \sqrt{- g} \left[ 4 - 10 \gamma \varphi \(\cos^2 \xi - \frac 1 5 \sin^2 \xi\) \right]\,,\label{WZ}
\ee
where $g$ denotes the determinant of the bulk metric. From (\ref{BI}) and (\ref{WZ}), we read off the sources of the bulk scalar field. 

We choose the bulk geometry to be $AdS_5$ in global coordinates (\ref{globalAdS}), so that $\sqrt{-\hat{g}} = r^3 [1+r^2/R_{AdS}^2]^{1/2}\sqrt{g_{S^3}}$ and $\sqrt{-g} = r^3\sqrt{g_{S^3}}$, and specialize to a spherical D3-brane of radius $R$ in $AdS_5$ that is localized at a point on the $S^5$. By $x$ we collectively denote the time coordinate and the coordinates on $S^3$. Due to the $SO(5)$ symmetry of the problem, the location of the brane on $S^5$ will only enter the action through the coordinate $\xi$ on $S^5$ (see \eq{xi}). We will be interested in D-branes near the conformal boundary of spacetime, in particular spherical branes with radius $R\gg R_{AdS}$, for which the sources $\mathcal J \equiv \frac{1}{\sqrt g} \frac{\delta S}{\delta \varphi} \Big {|}_{\varphi =0}$ for $\varphi$ reduce to
\be
\mathcal{J}_{DBI}(r) = 5\gamma \frac{\tau_3}{ R_{AdS}}  \(\cos^2 \xi - \frac 1 5 \sin^2 \xi\) r \delta(r-R)\,, \label{JDBI}
\ee
\ba
\mathcal{J}_{WZ} (r) = \left\{ \begin{array}{ll} \label{JWZ}
- 10 \gamma  \frac{\mu_3}{R_{AdS}}   \(\cos^2 \xi - \frac 1 5 \sin^2 \xi\)  & \textrm{$r \le R $}\\ 
0 & \textrm{$r  > R $}\,.
\end{array} \right. 
\ea

In fact, these expressions for the sources are valid not only in a pure $AdS$ background, but also for branes near the boundary of more general asymptotically $AdS$ backgrounds. Considering a static, spherically symmetric ansatz 
\be
ds_5^2 = - e^{- 2 \delta (r)} f(r) dt^2 + \frac{dr^2}{f(r)}  + r^2 d\Omega_3^2
\ee
and solving the equations of motion following from (\ref{5daction}),
\ba
\delta^{\prime}(r) &=&  -\frac 1 3 r \varphi^{\prime}(r)^2 \,,\\
r f^{\prime}(r) -2 + 2 f(r) &=& -\frac1 3 r^2 \left[ f(r)\varphi^{\prime}(r)^2 + 2 V(\varphi)\right]\,,\\
f(r) \left[\varphi^{\prime}(r) + r \varphi^{\prime \prime}(r) \right] &=& r \frac{\partial V}{\partial \varphi} + \frac 2 3 \varphi^{\prime}(r) \( r^2 V(\phi) - 3\)\,,
\ea
we find the asymptotic behavior 
\ba
f(r) &\sim& 1 + \frac{r^2}{R^2_{AdS}}\,,\\ \label{f(r)}
\varphi(x,r) &\sim&  \frac{1}{r^2}  \left[\alpha(x) \ln \(r\mu\)+  \beta(x) \right]\,,\label{varphi}\\
\delta(x,r) &\sim& \frac 1 3 \alpha(x)^2 \frac{ \ln^2\( r\mu\)}{r^4}\label{delta}\,.
\ea
In the limit of large radial coordinate we are interested in, (\ref{delta}) will not affect the computation of the D3-brane effective potential, since it will contribute to subleading order in $1/r$.

After computing the effective potential for the D3-brane transverse coordinates, we will want to compare it with the deformation \eq{Osquared} of the dual SYM theory. For that purpose, it will be useful to relate $R$ and the $S^5$ angles to canonically normalized scalar fields. From \eq{BI}, we can see that, for $R\gg R_{AdS}$, the scalar fields
\be \label{Phi}
\phi_1 \equiv \sqrt{\tau_3}  R_{AdS} R \cos \xi, \ \ \ \phi_2 \equiv  \sqrt{\tau_3}  R_{AdS} R \, \sin\xi \cos \Omega_1,\ \ \ \ldots
\ee
have canonical kinetic term
\be
S_{kin} = -\frac 1 2\int d^4 \tilde{x}\, \partial_{\alpha} \phi_i \partial^{\alpha} \phi_i
\ee
in the coordinate system $\tilde x^{\alpha} = (\tilde t \equiv t / R_{AdS}, \Omega_i)$ with metric
\be\label{tildecoords}
d\tilde s^2 = - d\tilde t^2 + d \Omega_3^2\,.
\ee 
To make contact with $\mathcal {N} =4$ \textrm{SYM}, the fields $\phi_i$ of \eq{Phi} play the role of eigenvalues of the fields $\Phi_i$ in \eq{operator}: 
\be\label{contact}
\Phi_i = {\rm diag}(\phi_i,0,\ldots,0),\ \ \ \ i=1,\dots ,6\,.
\ee

\subsection{Propagator of the bulk scalar field}

We now turn to the computation of the bulk propagator for the scalar field (satisfying the boundary condition \eq{BCs}). To solve the scalar equation of motion following from (\ref{5daction}), we separate variables writing
\be
\varphi(x,r) = e^{-i \omega t} Y_{\ell,m}(\Omega) \Psi(r)\,,
\ee
where $Y_{\ell}$ (with $\ell\ge 0$) is the $\ell^{th}$ spherical harmonics on $S^3$, satisfying
\be
\nabla^2_{S^3} Y_{\ell} = - \ell(\ell + 2) Y_{\ell}\,.
\ee
Letting $a = 1+ \frac 1 2(\ell+\omega)$ and $b=1+ \frac 1 2(\ell- \omega)$, and performing the change of coordinates 
\be
V = {r^2\over R^2_{AdS} + r^2},
\ee 
the propagator is constructed from the following two radial solutions \cite{Balasubramanian:1998sn}. The first solution,
\be
\Psi_1(V) = (1-V) V^{\ell/2}  {}_2 F_1(a,b,a+b; V)\, ,
\ee
is regular at the origin.\footnote{Other solutions only solve the equation of motion up to delta-function terms at the origin.} The second solution,
\ba
\Psi_2(V)& =& (1-V)V^{\ell/2} \big\{ {}_2F_1(a,b,1;1-V)\left[ 1+ C_{\infty} \ln (1-V)\right]+\\
&& C_{\infty} \sum_{k=1}^{\infty} (1-V)^k \frac{(a)_k(b)_k }{(k!)^2} [ \psi(a+k) + \psi(b+k) - 2 \psi(1+k) -\psi(a) - \psi(b) - 2\gamma_E] \big\}, \nonumber
\ea
where $\gamma_E$ denotes Euler's constant, satisfies the boundary conditions (\ref{BCs}) defined with the scale $\mu = 1/R_{AdS}$, provided that we choose 
\be\label{Cinf}
C_{\infty} =-  \frac f 2.
\ee
Combining the two expressions with the appropriate normalization factor, we obtain the Feynman propagator
\ba\label{prop}
G_f (x,V;x^{\prime},V^{\prime})& = &- \frac{\kappa_{10}^2}{R^2_{AdS} V_{S^5}} \int_{-\infty}^{\infty} \frac{d\omega}{2\pi} \sum_{\ell,m} \frac{\Gamma(a) \Gamma(b)}{\ell! (2\ell+2)}  \frac{1}{1- C_{\infty} [\psi(a) +\psi(b) +2 \gamma_E]}   \times \\ 
&& e^{-i \omega(t-t^{\prime})} Y_{\ell, m}(\Omega) Y_{\ell,m}(\Omega^{\prime}) \left[ \theta(V^{\prime}- V ) \Psi_1 (V) \Psi_2(V^{\prime}) + \theta(V- V^{\prime}) (V \leftrightarrow V^{\prime})\right] \,.\nonumber
\ea
\subsection{D3-brane effective potential}\label{D3potential}

Consider a probe D3-brane extended along a three-sphere with radius $R$ in global $AdS_5$ and localized at a point in $S^5$. In perturbation theory, the leading contribution to the D-brane effective potential is obtained by evaluating the D-brane action \eq{SD3} in the $AdS_5\times S^5$ background. Combining (\ref{BI}) and (\ref{WZ}) and making use of the BPS relationship $\tau_3 = \mu_3$, the leading order terms in the radial coordinate cancel among the two contributions. The term that survives in the DBI action results in an attractive quadratic potential $V\sim R^2$, corresponding to the conformal coupling of the massless scalar fields in the dual SYM theory on $\mathbb{R}\times S^3$ \cite{Seiberg:1999xz, Hubeny:2004cn}. This contribution is clearly independent of $f$, i.e., of the boundary condition on the bulk scalar field. (Note that $AdS_5\times S^5$ is compatible with the boundary conditions \eq{BCs} we consider. This was not the case for the boundary conditions studied in \cite{Hubeny:2004cn}.) 

The first contribution that is sensitive to the boundary condition is a diagram in which the brane emits and reabsorbes a $\varphi$ particle. This diagram involves a $\varphi$ propagator, which depends on $f$ according to \eq{prop} with \eq{Cinf}. Using this propagator and the sources (\ref{JDBI}), (\ref{JWZ}), we find the following term in the D3-brane effective action:
\ba
S_{eff} & = & \frac 1 2 \int d^4 x \int_0^R dr \sqrt{g} \left[\mathcal{J}_{BI} (r) + \mathcal{J}_{WZ} (r)\right]\times \nonumber  \\
&& \int d^4 x^{\prime}  \int_0^R dr^{\prime}\sqrt{g}\, G_f(x,r;x^{\prime},r^{\prime})  \left[\mathcal J_{BI} (r^{\prime}) + \mathcal J_{WZ} (r^{\prime})\right] \nonumber \\
&=&   f \frac {5} {12} \frac{\tau_3^2 \kappa_{10}^2}{V_{S^5}}  \int d^4 x R^4  \( \cos^2 \xi - \frac 1 5 \sin^2 \xi\)^2  \,.
\ea
Using the field redefinition \eq{Phi} and the change of variables that brings the boundary metric in the form \eq{tildecoords}, we can rewrite the effective potential as
\be\label{Veffprop}
\int d^4 \tilde x \,V_{eff} (\tilde x) = - f  \frac{5 \pi^2}{ 3 N^2} \int d^4 \tilde x\,  \left[ \phi_1^2 - \frac 1 5 \sum_{i=2}^6 \phi_i^2 \right]^2\,.
\ee 
For $f>0$, this is a quartic potential that pulls the spherical branes to the boundary of $AdS$. This potential agrees with what one would expect based on the dual ${\cal N}=4$ SYM theory. In particular, it is consistent with the ${\cal O}^2$ structure of \eq{Osquared} with \eq{operator}, as well as with the $N$-dependence of the deformation. In fact, the computation can be easily generalized to configurations with more than one brane. 

\begin{figure}
\begin{minipage}[t]{6.5cm}
\begin{center}
\includegraphics[width=6.5cm, height=6cm,clip]{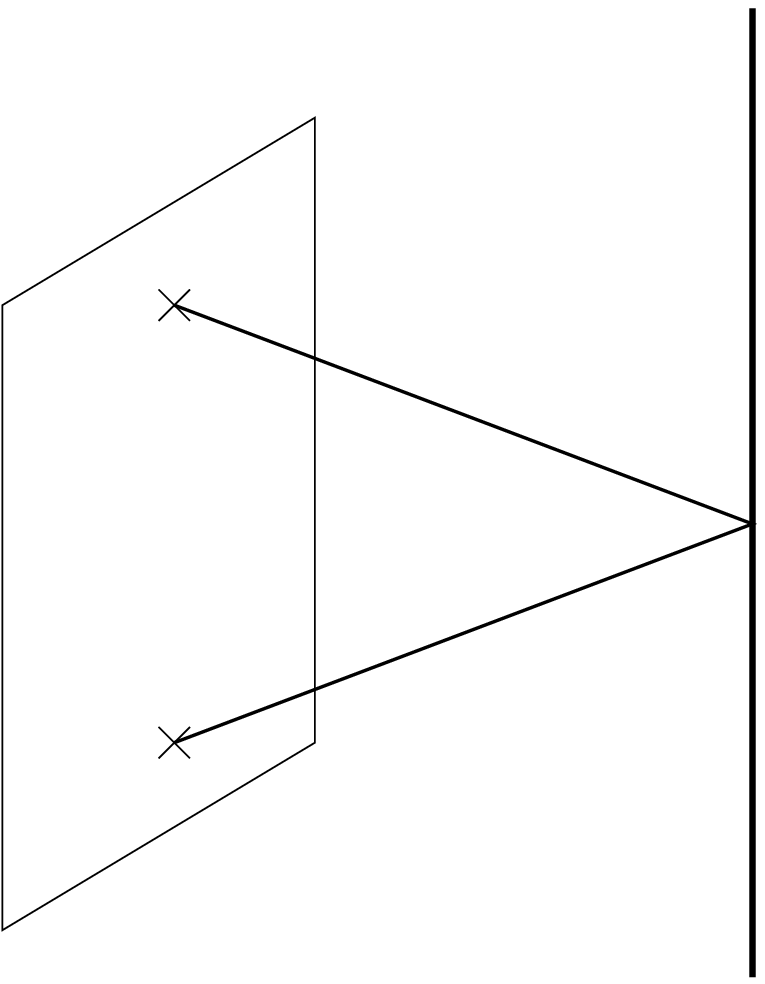}
\caption{\label{single}A virtual $\varphi$ particle is emitted by the D3-brane, interacts with the boundary at $r= \Lambda$ and is reabsorbed by the brane.}
\end{center}
\end{minipage}
\hfill
\begin{minipage}[t]{6.5cm}
\begin{center}
\includegraphics[width=6.5cm,height=6cm,clip]{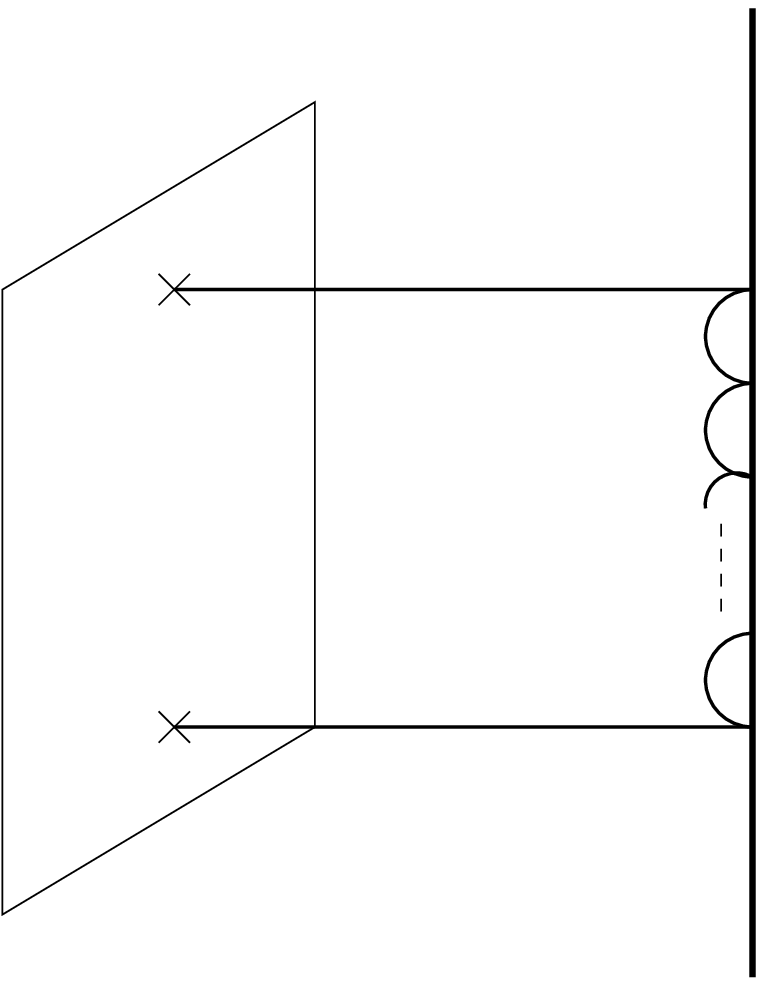}
\caption{\label{multiple} A virtual $\varphi$ particle emitted by the D3-brane interacts an arbitrary number of times with the boundary before reabsorption.}
\end{center}
\end{minipage}
\end{figure}

\begin{figure}
\begin{center}
\epsfig{file=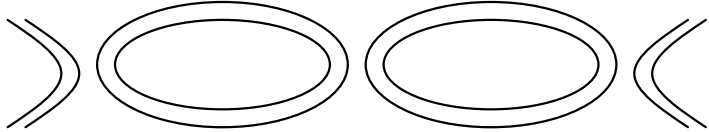, width=10cm}
\end{center}
\caption{A two-loop example of factorizable diagrams that survive in the large $N$ limit and renormalize the coupling $f$.}\label{figHigher}
\end{figure}

This is the main result of this section, and we could stop here. However, our derivation crucially relied on the fact that the boundary condition \eq{BCs} is linear, so that it could be fully taken into account by the modified propagator \eq{prop}. Equivalently, the boundary term \eq{bdry5d} is quadratic in $\varphi$, so that its effect can be absorbed in a modification of the propagator. This class of boundary conditions is very special. In fact, in the next section we will be interested in a non-linear boundary condition corresponding to a cubic boundary term and a triple trace interaction in the dual field theory. Therefore, we will now compute the D3-brane potential in a way that easily generalizes to non-linear boundary conditions. The idea is to work with the standard $f=0$ propagator (corresponding to supersymmetric boundary conditions) and to treat the $f$-dependent boundary term in \eq{bdry5d} as an interaction. Then, as illustrated in figure~\ref{single}, a virtual $\varphi$ particle emitted by a D3-brane can propagate to the boundary and ``feel'' the $f$-dependent boundary interaction before being reabsorbed by the D3-brane (the effect of the $f$-independent boundary term is already accounted for in the $f=0$ propagator). In fact, the virtual $\varphi$ particle can interact with the boundary an arbitrary number of times before being reabsorbed (see figure~\ref{multiple}). 
Our previous computation, where the effect of the modified boundary condition was incorporated in a modified propagator, amounts to a resummation of all these contributions (which is possible for linear boundary conditions but not in more general cases). 
Let us thus compute the contribution to the D3-brane effective potential from a virtual $\varphi$ particle interacting with the boundary a single time. Using the $f=0$ propagator \eq{prop} (with $C_\infty=0$) and the expressions (\ref{JDBI}), (\ref{JWZ}) for the sources, we find
\ba\label{Seffpert}
S_{eff} &=& \frac{f}{2\(1+f \ln \frac{\Lambda}{R_{AdS}} \)} \frac{V_{S^5}}{ R_{AdS} \kappa_{10}^2} \int_{\partial} d^4 x   \sqrt{g_{bdy}} \int d^5 x^{\prime} \sqrt{g} \left[\mathcal J_{BI} (r^{\prime}) + \mathcal J_{WZ}(r^{\prime})\right]   \times\nonumber \\ 
&& G_{f=0} (x^{\prime},r^{\prime};x ,\Lambda)\int d^5 x^{\prime \prime} \sqrt{g} G_{f=0}(x ,\Lambda;x^{\prime \prime},r^{\prime \prime}) \left[\mathcal J_{BI} (r^{\prime \prime}) + \mathcal J_{WZ}(r^{\prime \prime})\right] \,,
\ea
which becomes
\be
\int d^4 \tilde x\, V_{eff} (\tilde x) = -\frac {f}{1+ f \ln \frac{\Lambda}{R_{AdS}}}  \frac{5 \pi^2}{ 3 N^2} \int d^4 \tilde x\, \left[ \phi_1^2 - \frac 1 5 \sum_{i=2}^6 \phi_i^2 \right]^2\,.
\label{Veffpert}
\ee
The difference between \eq{Veffprop} and \eq{Veffpert} is that $f$ got replaced with $f/[1+f\ln(\Lambda/R_{AdS})]$ (which formally vanishes when the cutoff $\Lambda$ is removed). From the point of view of our computations, the difference corresponds to the diagrams in figure~\ref{multiple}, with multiple boundary interactions -- taking them into account will convert $f/[1+f\ln(\Lambda/R_{AdS})]$ into $f$. From a dual field theory point of view, the difference lies in the scale at which the couplings are defined (cf.\ \cite{Witten:2001ua}). When expressing the asymptotic behavior \eq{scalar} of the scalar field, we had to choose a scale $\mu$ to define the logarithm $\ln \(r\mu\)$, and we chose%
\footnote{
If in \eq{scalar} we had left a generic scale  $\mu$, \eq{Veffprop} would have read 
\be\label{scalemu}
\int d^4 \tilde x\, V_{eff} (\tilde x) = - \frac{f}{1- f \ln \(\mu R_{AdS}\)}  \frac{5 \pi^2}{ 3 N^2} \int d^4 \tilde x\, \left[ \phi_1^2 - \frac 1 5 \sum_{i=2}^6 \phi_i^2 \right]^2 \,.
\ee
} 
$\mu=1/R_{AdS}$, the scale appearing in the metric \eq{globalAdS}. This scale corresponds to a renormalization scale $1/(\mu R_{AdS}^2)=1/R_{AdS}$ in the boundary theory \cite{Witten:2001ua}. On the other hand, $f/[1+f\ln(\Lambda/R_{AdS})]$ is the coupling defined at the UV cutoff scale $\Lambda/R_{AdS}^2$ of the dual field theory. From a large $N$ field theory perspective, the relation between the couplings at both scales is given by a resummation of (factorizable) planar diagrams with an arbitrary number of loops -- the two-loop diagram is drawn in figure~\ref{figHigher}.  
\subsection{Expanding D3-branes, five-form flux and geometric transition}

We have seen that the radius $R$ of a spherical D3-brane in $AdS_5$ with modified boundary condition (labeled by $f$) on a quadrupole deformation mode of $S^5$ feels a quartic potential. For $f>0$, this potential tries to blow up the D3-brane to infinite radius in finite time. For sufficiently small spherical D3-branes, this is prevented by the positive $R^2$ term in the potential, corresponding to the conformal coupling of the SYM scalars to the curvature of $S^3$. For sufficiently large branes, the quartic potential wins and the branes are pushed to the boundary of $AdS_5$ in finite time.

In which contexts do such large spherical branes play a role? On the one hand, one could start with a system without them and they could be spontaneously created by quantum tunneling. This could happen, for instance, to the pure $AdS_5\times S^5$ with modified ($f>0$) boundary conditions, which is known to be only meta-stable (we expect the nucleating spherical branes to be closely related to the instanton solutions of \cite{Hertog:2004rz}). This is obviously a dynamical process that cannot be described in classical supergravity. On the other hand, one could consider an initial state with large spherical D3-branes present and study the time evolution of this state. This is analogous to the point of view taken in \cite{Hertog:2004rz, Craps:2007ch}, where the interest was in the evolution into a big crunch, not so much in instabilities of pure $AdS_5$.

To make the analogy between spherical D3-brane evolution and the (super)gravity solutions of \cite{Hertog:2004rz, Craps:2007ch} more precise, we have to relate the radius $R$ of spherical D3-branes to the scalar field $\varphi$ appearing in \eq{5daction}. As we shall discuss in section~2.5.1, this was done in \cite{Freedman:1999gk}, at least for the related system of flat D3-branes and $AdS_5$ in Poincar\'e coordinates. The upshot is that configurations that fit in the consistent truncation \eq{5daction} correspond not to a single D3-brane but to specific distributions of D3-branes, and thus to specific distributions of radii. So to make contact with the supergravity solutions of \cite{Hertog:2004rz, Craps:2007ch}, we should start with such a distribution of large spherical D3-branes.

One point that may appear puzzling at first is related to the five-form flux through the $S^5$ in the ten-dimensional supergravity solutions of \cite{Craps:2007ch}. A spherical D3-brane acts as a domain wall, with the five-form flux inside being one unit smaller than the flux outside. If the big crunch instability in the solutions of \cite{Craps:2007ch} corresponds to spherical D3-branes expanding to infinite size, one might thus expect that at any value of the radial coordinate $r$, the five-form flux should decrease as a function of time as spherical D3-branes expand from radius smaller than $r$ to radius bigger than $r$. However, if we compute the flux through $S^5$ for the solutions of \cite{Craps:2007ch}, we find that the flux remains constant as the bulk scalar field evolves in time:
\be
\int_{S^5} \hat{G}_5 = - \int d \xi d\Omega_5 U \Delta^{-2} R^4_{AdS} = 16 \pi^4 g_s \alpha^{\prime 2} N\,.
\ee
A related point is that the solutions of \cite{Craps:2007ch} solve the supergravity equations of motion without any D3-brane sources present. The resolution of this paradox lies in the concept of geometric transition \cite{Vafa:2000wi, Lin:2004nb}, which relates a situation with D-brane sources explicitly present to a situation with the D-branes replaced by flux (and the location of the D-branes ``cut out'' of the space).\footnote{Of course, in the present situation, neither description is valid in the regime in which the space-time is highly curved.} In our spherical D3-brane picture, we considered the shape of the $S^5$ not to change with time and treated the D3-branes as sources; in the solutions of \cite{Craps:2007ch}, the D3-branes are not explicitly present, but the shape of the $S^5$ changes in such a way that the would-be locations of the D3-branes are always ``cut out'' of the spacetime.

\subsubsection{Relating D3-brane positions with the bulk scalar field}

As mentioned earlier in this section, the bulk scalar field $\varphi$ can be related to D3-brane positions. The near-horizon limit of a distribution $\sigma$ of parallel D3-branes in $n$ transverse dimensions is given by 
\ba
ds^2& =& \frac{1}{\sqrt{H}}\left( - dt^2 + dx_1^2 +dx^2_2 +dx_3^2 \right) + \sqrt{H} \sum_{i=1}^6 dy_i^2, \nonumber\\
H &=& \int d^n \omega \sigma(\vec{\omega}) \frac{R_{AdS}^4}{|\vec{y} -\vec{ \omega}|^4} \,.
\ea
For generic distributions $\sigma$, this does not fall in the class of metrics \eq{10dg}, so generic D3-brane configurations cannot be described using the consistent truncation \eq{5daction}. In \cite{Freedman:1999gk}, it was shown that specific configurations of D3-branes do have near-horizon geometries that can be described using the consistent truncation. For instance, an $SO(5)$ symmetric configuration of D3-branes distributed on a one-dimensional interval of length $\ell$ according to $\sigma(\vec{\omega}) = \frac{2}{\pi \ell^2} \sqrt{\ell^2 - |\vec{\omega}|^2}$ gives rise to the metric
\be
ds^2 = \frac{\xi r^2}{\lambda^3R_{AdS}^2}\left[ dx_{\mu}^2 + \frac{R_{AdS}^4}{r^4} \frac{dr^2}{\lambda^6}\right] + \frac{\lambda^3R_{AdS}^2}{\xi}\left[ \xi^2 d\theta^2 + \cos^2 \theta d\Omega_4^2\right]\,,
\ee
with 
\be
\lambda^{12} = 1 +\frac{\ell^2}{r^2}\,, \qquad  \xi^2 = 1+ \frac{\ell^2}{r^2} \cos^2 \theta\,.
\ee
Comparing with (\ref{10dg}), one finds that in the regime of large radial coordinate, the scalar field profile is
\be
\varphi(x,r) = \frac{\ell^2}{6 \gamma} \frac{1}{r^2}\,,
\ee
with $\gamma$ defined after \eq{5daction}. Therefore, the coefficient of the asymptotic fall-off of $\varphi$ is directly related to the size over which D3-branes are spread out. This conclusion holds for flat D3-branes and in the Poincar\'e coordinate system, which appears when taking near-horizon limits. However, since near the boundary of $AdS$ spherical branes are almost flat and Poincar\'e coordinates are a good approximation to global coordinates, we expect the conclusion to extend to large spherical D3-branes in global $AdS$. 

\setcounter{equation}{0}

\section{M2-branes in $AdS_4 \times S^7/\mathbb{Z}_k$}

In this section, we study spherical M2-branes in global $AdS_4 \times  S^7$ and $AdS_4 \times S^7/\mathbb{Z}_k$. M-theory allows a consistent truncation to four-dimensional gravity with a negative cosmological constant coupled to a single scalar field \cite{Duff:1999gh}. This scalar corresponds to a quadrupole deformation of the seven-sphere, has a mass that is above the Breitenlohner-Freedman bound \cite{Breitenlohner82} and preserves a subgroup of the full $SO(8)$ symmetry. Along the lines of the discussion in section 2, we compute the M2-brane effective potential as a function of the boundary conditions on this bulk scalar field. Therefore, we determine the coupling of the five-dimensional scalar to M2-branes considering the M2-brane action and the consistent truncation ansatz that relates the eleven and four-dimensional solutions. We observe that the modified boundary conditions correspond to adding a cubic boundary interaction to the bulk action and we compute the interaction of the M2-brane with the boundary via this cubic vertex. (Note that, due to the non-linearity of the boundary conditions we will consider, their effect cannot be absorbed in a modification of the scalar field propagator.) 

Specifically, we will consider a class of $AdS$ invariant boundary conditions that corresponds to adding a marginal triple trace deformation to the dual field theory \cite{Hertog:2004rz, Aharony:2008ug, Craps:2009qc}. We will first discuss M-theory on $AdS_4 \times S^7$, which is obtained as the near-horizon geometry of M2-branes in flat space and is dual to the $k=1$ case of ABJM theory. Then we will consider ABJM theory for general $k$, which corresponds to M2-branes on a $\mathbb Z_k$ orbifold of $\mathbb{C}^4 $, which have $AdS_4 \times S^7/\mathbb{Z}_k$ as near-horizon geometry. 

In section 3.1 we review the bulk setup as well as some relevant aspects of ABJM theory. We identify the deformation that corresponds, according to the $AdS$/CFT dictionary, to our choice of boundary conditions. In section 3.2, we use the lift to the eleven-dimensional solution to identify the coupling of the bulk scalar field to spherical M2-branes. In section 3.3 we compute the propagator for the four-dimensional scalar field (for standard boundary conditions). In section 3.4, we compute the potential for spherical M2-branes in $AdS_4 \times S^7$. Finally, in section 3.5, we extend the discussion to M2-branes in $AdS_4 \times S^7/\mathbb{Z}_k$ and comment on the 't~Hooft limit of the result. 

\subsection{Setup}
M-theory in asymptotically $AdS_4 \times S^7$ spacetimes has four-dimensional $SO(8)$ gauged $\mathcal N =8$ supergravity as its low energy limit. This theory allows a consistent truncation to four-dimensional gravity coupled to a single scalar field that preserves an $SO(4) \times SO(4)$ symmetry
\be\label{S4d}
S = \frac{V_{S^7}}{\kappa_{11}^2} \int d^4 x \sqrt{g} \left[ \frac{R}{2} - \frac  1 2 \partial_{\mu} \varphi \partial^{\mu} \varphi +\frac{1}{R_{AdS}^2}\( 2+ \cosh (\sqrt{2} \varphi) \)\right]\,,
\ee
with $2 \kappa_{11}^2 = (2\pi)^8 \ell_p^9$ in terms of the eleven-dimensional Planck length and $V_{S^7} = \pi^4 (2 R_{AdS})^7 / 3 $. The potential has a maximum for vanishing scalar field that corresponds to the $AdS_4$ vacuum solution. Small fluctuations around the the $AdS$ solution have a mass $m^2 = -2 /R^2_{AdS}$, which is above the BF bound (see footnote \ref{MBF}), and therefore the maximally supersymmetric solution, with the standard boundary conditions, is both perturbatively and  non-perturbatively stable. 
In global coordinates the $AdS_4$ metric reads
\be
ds^2 = -\(1+ \frac{r^2}{R^2_{AdS}}\) dt^2 + \frac{dr^2}{1+\frac{r^2}{R_{AdS}^2}} + r^2 d\Omega_2^2\,.\label{globalAdS4}
\ee
In any asymptotically $AdS$ solution, the scalar field behavior at large radial coordinate is
\be\label{varphi4}
\varphi(x,r) = \frac{\alpha(x)}{r} +\frac{\beta(x)}{r^2}\,,
\ee
where $x$ collectively denotes the time coordinate and the $S^2$ angles. The usual boundary conditions correspond to taking either $\alpha=0$ (which can be chosen for any $m^2$) or $\beta=0$ (which can be chosen for scalars in the mass range $m_{BF}^2 < m^2 < m^2_{BF} +1/R^2_{AdS}$). There exists however a whole one-parameter family of $AdS$ invariant boundary conditions, i.e., boundary conditions that preserve the asymptotic symmetries of $AdS$ spacetime, which allow the construction of well-defined and finite Hamiltonian generators \cite{Hertog:2004dr, Henneaux:2006hk}. The general class is 
\be\label{BCs4}
\beta(x) = - h \alpha(x)^2\,,
\ee
where $h$ is an arbitrary constant \cite{Hertog:2004dr}. For $h \neq 0 $, smooth asymptotically $AdS$ initial data can evolve into a big crunch singularity \cite{Hertog:2004rz}.\footnote{The sign of $h$ is irrelevant, as it can be changed by redefining $\varphi\rightarrow -\varphi$.}

Adding to the bulk action \eq{S4d} the boundary term 
\be\label{Sbdy4d}
S_{bdy}= \frac{V_{S^7}}{\kappa_{11}^2 R_{AdS}} \int_{\partial} d^3 x \sqrt{g_{bdy}} \(-\frac 1 2 \varphi^2 + \frac h 3 \varphi^3 + \frac{h^2}{2} \varphi^4 \)\,,
\ee
the boundary condition \eq{BCs4} follows from a variational principle. As in section \ref{setup5d}, we have introduced a regularized boundary $\partial$ in spacetime, located at $r= \Lambda$. 

M-theory in asymptotically $AdS_4\times S^7$ spacetimes with $\beta =0$ boundary conditions is dual to the three-dimensional superconformal field theory that describes the low energy dynamics of coincident M2-branes. In \cite{Aharony:2008ug}, Aharony, Bergman, Jafferis and Maldacena (ABJM) proposed a specific three-dimensional $\mathcal N=6$ superconformal $U(N) \times U(N)$ Chern-Simons-matter theory with levels $k$ and $-k$ as the world-volume theory of $N$ coincident M2-branes on a $\mathbb{C}^4 / \mathbb{Z}_k$ singularity. Besides the two $U(N)$ gauge fields $A$ and $\hat A$, the theory contains scalar fields $Y^A$, $A=1,\dots,4$, transforming in the fundamental representation of the $SU(4)_R$ R-symmetry group and in the bifundamental $(N, \bar N)$ of the gauge group. The Hermitean conjugate scalar fields $Y^{A\dagger}$ transform in the anti-fundamental representation of $SU(4)_R$ and in the $(\bar N, N)$ of the gauge group. The action reads
\ba \label{ABJM}
S_0 &= &\int d^3 x \Big[ \frac{k}{4 \pi} \e^{abc} \Tr \(A_a \partial_b A_c +\frac{2i}{3} A_a A_b A_c -  \hat A_a \partial_b \hat A_c - \frac{2i}{3} \hat A_a \hat A_b \hat A_c\)\nonumber\\
&&- \Tr (D_a Y^A)^{\dagger} D^a Y^A + V_{bos} + \mbox{terms with fermions} \Big]\,,
\ea
where $V_{bos}$ is a sextic potential for the scalars and we will not need the fermion fields explicitly in the following. The bulk setup we considered in this section corresponds to the case $k=1$, for which the transverse space to the M2-branes is simply $\mathbb{R}^8$. We will discuss Chern-Simons level $k>1$ in section \ref{D2}. In ABJM theory on $\mathbb{R}\times S^2$, the four complex scalars $Y^A$ effectively get mass $m^2 = 1/4$ due to the conformal coupling to the curvature of the $S^2$ (which we choose to have unit radius; see footnote \ref{confcoupling}). The boundary condition $\beta = -h \alpha^2$ corresponds to adding a marginal triple trace deformation to the boundary action 
\be\label{hdef}
S = S_0 + \frac h 3 \int d^3 x \mathcal O^3 \,.
\ee
Here, $\mathcal O$ is the dimension one chiral primary operator
\be
\mathcal O = c \,Tr \(Y^1 Y_1^{\dagger} + Y^2Y_2^{\dagger} - Y^3 Y_3^{\dagger} - Y^4Y_4^{\dagger}\), \label{O4}
\ee
which preserves the same $SO(4) \times SO(4)$ subgroup of $SO(8)$ as the bulk scalar $\varphi$ in the consistent truncation. The constant $c$ in \eq{O4} depends on the two dimensionless parameters $N$ and $k$ in a way that we will determine in section \ref{Mpotential}. 

Taking the decoupling limit of a system of coincident M2-branes in eleven-dimensional flat space, we observe that the world-volume field theory of ABJM with $k=1$ has a dual gravitational description in terms of M-theory on $AdS_4 \times S^7$. In this description, the eigenvalues of the four complex scalar fields $Y^A$ and of their Hermitean conjugates correspond to M-brane positions in the transversal space as in the case of $\mathcal N =4$ SYM (see section 2.3 of \cite{Aharony:2008ug} for a discussion, and \cite{Mukhi:2008ux} for more details). The deformation \eq{hdef} provides a sextic potential for these positions that is unbounded below and above, whatever the sign of $h$. This potential is sufficiently strong to make the eigenvalues become infinite in finite time, corresponding to M2-branes reaching the conformal boundary of $AdS$ in finite time. In sections~3.4 and~3.5, we will obtain the effective potential of spherical M2-branes as a function of the boundary conditions in the bulk and show that it matches the deformation \eq{hdef}.  
\subsection{Coupling of the bulk scalar field to spherical M2-branes}

The lift of the four-dimensional solution of \eq{S4d} to eleven-dimensional supergravity is given in \cite{Cvetic:1999xp}. Letting $F = e^{\varphi /\sqrt 2 }$ and  $ \tilde{ \Delta} = F \cos^2 \theta + F^{-1} \sin^2 \theta$, the full eleven-dimensional metric and four-form read
\ba
ds_{11}^2 &=&\tilde{ \Delta}^{2/3} ds_4^2 + 4 R^2_{AdS} \left[\tilde{ \Delta}^{2/3}d\theta^2 + \tilde{ \Delta}^{-1/3} \( F \sin^2\theta d\Omega_3^2 + F^{-1} \cos^2 \theta d\tilde{\Omega}_3^2 \) \right] \,,\label{11metric}\\
\hat F_4& =&-  \frac{U}{R_{AdS}} \e_4 + 8 R_{AdS} \sin \theta \cos \theta F^{-1} \ast dF \wedge d\theta\,, \label{F4}
\ea
with
\be
U = -2-F^2 \cos^2 \theta - F^{-2} \sin^2 \theta\,.
\ee
We have chosen coordinates in terms of which the unit seven-sphere metric would read
\be \label{sevensphere}
d\Omega_7^2  = d\theta^2 + \sin^2 \theta  d\Omega_3^2 + \cos^2 \theta  d\tilde{\Omega}_3^2\,,
\ee 
with $0 \le \theta \le \pi$ and, in \eq{F4}, $\epsilon_4$ and $\ast$ are the four-dimensional volume-form and dual. 

We want to repeat the procedure we carried out in section \ref{coupling} and consider a probe M2-brane in the eleven-dimensional lifted solution to determine its coupling to the bulk field $\varphi$. The action of the probe brane is 
\be\label{SM2}
S_{M2} = S_{DBI}+S_{WZ} = -\tau_2 \int d^3 x \sqrt{ \hat G} + \mu_2 \int \hat C_3\,,
\ee
where $\hat G$ is the determinant of the pull-back of the eleven-dimensional metric to the M2-brane worlvolume, $d\hat C_3 = \hat F_4$ and where, for convenience, we have split the M2-brane action in analogy with the conventional notation for D-brane actions . The tension and charge are $\tau_2 = \mu_2 = 2\pi (2\pi \ell_p)^{-3}$.
In the static gauge and to linear order in $\varphi$, the ``DBI'' part of the action reads
\be \label{SDBI4}
S_{DBI} = -\tau_2 \int d^3 x \sqrt{-\hat{g}}\left[ 1 + \frac{1}{\sqrt 2} \varphi \(\cos^2 \theta -\sin^2 \theta\) + \frac 1 2 g_{ij} \partial_a x^i \partial^a x^j \right]\,,
\ee
where $\hat g$ is the determinant of the pull-back of the four-dimensional metric $g_{\mu\nu}$ to the three-dimensional world-volume, the index $a$ labels the coordinates along the M2-brane world-volume and the index $i$ runs over the eight transverse directions. The Wess-Zumino action is
\be\label{SWZ4}
S_{WZ} =\frac{ \mu_2}{R_{AdS}} \int_{V_4} d^4 x \sqrt {-g} \left[ 3 + \sqrt 2 \varphi   \(\cos^2 \theta -\sin^2 \theta\)  \right]\,,
\ee
expressed as an integral over the four-dimensional volume enclosed by the M2-brane. Here $g$ denotes the determinant of the bulk metric.

Choosing the bulk geometry to be $AdS_4$ in the global coordinates \eq{globalAdS4} and specializing to a spherical M2-brane of radius $R$ that is localized on $S^7$, the sources for the scalar field $\varphi$ are
\be
\mathcal{J}_{DBI}(r) = - \frac{\tau_2}{R_{AdS}} \frac{1}{\sqrt 2} (\cos^2 \theta -  \sin^2 \theta) r \delta(r-R)\,, \label{JDBI4}
\ee
\ba
\mathcal{J}_{WZ} (r) = \left\{ \begin{array}{ll} \label{JWZ4}
2  \frac{\mu_2 }{R_{AdS}}\frac{1}{\sqrt 2} (\cos^2 \theta - \sin^2 \theta)  & \textrm{$r \le R $}\\ 
0 & \textrm{$r  > R $}\,.
\end{array} \right. 
\ea
To interpret the result we will obtain for the effective potential in the language of ABJM theory and compare it with the deformation \eq{hdef}, we introduce canonically normalized scalars on the conformal boundary of metric $d\tilde s^2 = -d \tilde t^2 + d\Omega_3^2$. The relation between these scalars and the M2-brane radius $R$ and $S^7$ angles is
\begin{eqnarray}\label{Phi4d}
\phi_1 &\equiv& 2 R_{AdS} \sqrt{\tau_2 R} \cos \theta  \cos \Omega_1, \  \ \ \phi_2 \equiv 2 R_{AdS} \sqrt{\tau_2 R} \cos \theta  \sin \Omega_1 \cos \Omega_2, \ \ \dots \nonumber \\ 
\phi_5 &\equiv& 2 R_{AdS} \sqrt{\tau_2 R} \sin \theta  \cos \Omega_4, \ \ \  \phi_6 \equiv 2 R_{AdS} \sqrt{\tau_2 R} \sin \theta  \sin \Omega_4 \cos \Omega_5, \ \ \dots
\end{eqnarray}
as can be seen form \eq{SDBI4}.  Complex combinations of these fields will correspond to eigenvalues of the fields $Y^A$ appearing in \eq{ABJM}.
\subsection{Propagator of the bulk scalar field}

To compute the propagator for the field $\varphi$, we follow again \cite{Balasubramanian:1998sn} and separate variables as
\be
\varphi(x,r) = e^{-i \omega t} Y_{\ell, m} (\Omega) \Psi(r)\,,
\ee
where the spherical harmonics satisfy $\nabla^2_{S^2} Y_{\ell} = - \ell(\ell + 1) Y_{\ell}$, with $\ell \ge 0$. The radial solution that is regular in the interior is
\be \label{solint4}
\Psi_1(V) = (1-V) V^{\ell/2}\, {}_2F_1\(a,b,a+b-\frac 1 2 ; V\)\,.
\ee
The propagator is constructed from \eq{solint4} and the radial solution with asymptotic behavior
\be\label{solinf4}
\Psi_2 (V) = (1-V)^{1/2} V^{\ell/2}\left[  {}_2F_1\(a-\frac 1 2 ,b -\frac 1 2 , \frac 1 2 ;1-  V\) + K_{\infty} \, {}_2F_1\(a,b, \frac 3 2 ;1- V\)\right] \,, 
\ee
in terms of a coefficient $K_{\infty}$ that implements the specific choice of boundary conditions. The standard supersymmetric choice $\beta=0$ sets $K_{\infty} = 0$. In \eq{solint4} and \eq{solinf4}, we have again denoted $a = 1+ \frac 1 2(\ell+\omega)$, $b=1+ \frac 1 2(\ell- \omega) $ and $V = r^2/(R^2_{AdS} + r^2)$. 

Combining the two solutions above with the appropriate normalization factor, we obtain the Feynman propagator
\ba\label{prop4}
G_F (x,V;x^{\prime},V^{\prime})& = &-\frac{\kappa_{11}^2}{R_{AdS} V_{S^7}} \int_{-\infty}^{\infty} \frac{d\omega}{2\pi} \sum_{\ell,m} \frac{\Gamma(a-\frac 1 2) \Gamma(b-\frac  1 2)}{2 \sqrt{\pi} \Gamma(a+b -\frac 1 2)} e^{-i \omega(t-t^{\prime})}\times \\ 
&& Y_{\ell, m}(\Omega) Y_{\ell,m}(\Omega^{\prime}) \left[ \theta(V^{\prime}- V ) \Psi_1 (V) \Psi_2(V^{\prime}) + \theta(V- V^{\prime}) (V \leftrightarrow V^{\prime})\right] \,.\nonumber
\ea

\subsection{M2-brane effective potential}\label{Mpotential}

In this section, we compute the effective potential for the radial coordinate $R$ of a probe M2-brane that extends along a two-sphere and is localized on $S^7$. We evaluate the M2-brane action \eq{SM2} in an $AdS_4 \times S^7$ background.\footnote{Actually the requirement of an asymptotically $AdS_4 \times S^7$ background is sufficient.} The BPS relation between the charge and tension of the brane guarantees the cancellation of the leading order terms in the radial coordinate, as can be seen from \eq{SDBI4} and \eq{SWZ4}. The $h$-independent term that survives the cancellation is an attractive potential linear in $R$, which corresponds to the conformal coupling of the scalar fields $Y^A$ of the dual theory on $\mathbb R \times S^2$. The dependence of the potential on the modified boundary conditions shows up, to lowest order, in a Feynman diagram in which the probe brane interacts with the boundary, exchanging scalar $\varphi$ modes through the cubic coupling in \eq{Sbdy4d}. Generalizing (\ref{Seffpert}) to a cubic boundary interaction, we obtain
\be\label{Seffpert4}
S_{eff} = \frac h 3 \frac{\tau_2^3 \kappa_{11}^4 R_{AdS}}{V_{S^7}^2}\int d^3 x R^3 \frac{1}{2\sqrt 2}\(\cos^2 \theta - \sin^2 \theta\)^3\,.
\ee
In terms of the boundary fields of equation \eq{Phi4d} and of the background metric $\tilde g$ defined above \eq{Phi4d}, the result becomes
\be\label{Veffpert4}
\int d^4 \tilde x\, V_{eff}(\tilde x) = -\frac {h}{N^3} \frac{3 \pi^2}{8} \int d^4 \tilde x\, \left[ \sum_{i=1}^4 \phi_i^2 - \sum_{i=5}^8 \phi_i^2 \right]^3 \,. 
\ee
In the last step we have used the relation $2R_{AdS} /\ell_p = (2^5 \pi^2 N)^{1/6}$ \cite{Aharony:2008ug}, which relates the radius of $AdS$ with $N$ units of flux to the eleven-dimensional Planck length. For a non-vanishing value of $h$, this is a sextic potential with unstable directions. Using a similar argument as after \eq{Veffprop}, one can see that the potential \eq{Veffpert4} matches the deformation \eq{hdef} of ABJM theory. For $k=1$, it fixes the $N$-dependence of the operator $\mathcal O$ in \eq{O4} to be $c\sim 1/N$.

\subsection{Extension to $k>1$}\label{D2}

The result of the previous section corresponds to the $k=1$ case of ABJM theory.  To generalize to arbitrary $k$, consider the $\mathbb{Z}_k$ orbifold of the eleven-dimensional supergravity solution. In \cite{Nilsson:1984bj, Aharony:2008ug}, the metric on the seven-sphere was written in a Hopf-fibered way: 
\be\label{S7fibered}
ds^2_{S^7} = (d \chi + \omega)^2 + ds^2_{\mathbb{C}P^3}
\ee
with $\chi$ periodic with periodicity $2\pi$. The $\mathbb{Z}_k$ action simply changes the periodicity of the coordinate $\chi$ to $2\pi/k$. Since the volume of the quotient space is smaller by a factor $k$ than the original one, in order to have $N$ units of flux of the four-form \eq{F4} on the quotient space, we need to start with $N^{\prime} = k N$ units on the covering space. The circle labeled by $\chi$ can be interpreted as the M-theory circle.

In the parametrization \eq{sevensphere} of $S^7$, we can exhibit the $\chi$ direction by writing 
\be
d\Omega_3^2 =  \left[d(\chi + \tilde\chi) + \omega\right]^2 + ds^2_{\mathbb{C}P^1}\,, \quad  d\tilde{\Omega}_3^2 = \left[d(\chi - \tilde\chi) + \tilde{\omega} \right]^2 + d\tilde s^2_{\mathbb{C}P^1}\,.
\ee
where as before $\chi$ has periodicity $2\pi/k$ after the $\mathbb{Z}_k$ identification.

The $\mathbb{Z}_k$ identification on the $\chi$ direction rescales the volume of the $S^7$, $V_{S^7}$, by a factor $1/k$. As pointed out in \cite{Craps:2009qc}, the bulk scalar field $\varphi$ survives the $\mathbb Z_k$ quotient, so, to extend the previous discussion to an arbitrary value of the Chern-Simons level, it suffices to trace back its contributions in the computation of the effective potential. As a consequence of the orbifolding, the actions \eq{S4d} and \eq{Sbdy4d} get rescaled by a factor $1/k$ and therefore, the propagator for the field $\varphi$ \eq{prop4} has to be multiplied by a factor $k$. The M2-brane action \eq{SM2} is unaffected by the identification since the M-theory direction is transverse to the M2-brane. The overall effect of the $\mathbb Z_k$ action is to rescale the final result \eq{Veffpert4} by a factor $k^2$. 
Substituting $N^{\prime} = k N$, it combines into
\be\label{Veffpert4k}
\int d^4 \tilde x\, V_{eff}(\tilde x) = -\frac {h}{ k N^3} \frac{3 \pi^2}{8} \int d^4 \tilde x \,\left[ \sum_{i=1}^4 \phi_i^2 - \sum_{i=5}^8 \phi_i^2 \right]^3\,.
\ee

We now discuss various parameter regimes of the theory of $N^{\prime}=kN$ M2-branes on a $\mathbb{C}^4 / \mathbb Z_k$ singularity to comment the $k$ and $N$ dependence of the effective potential. As discussed in \cite{Aharony:2008ug}, the radius of the M-theory circle in Planck units is of order $R_{AdS} /k \ell_p \sim (k N)^{1/6} / k$, while the radius of the $\mathbb{C} P^3$ factor is always large in Planck units if $k N \gg 1$. Thus the M-theory description reduces to a weakly coupled type IIA string theory whenever $k^5 \gg N$. In this limit, the M2-action \eq{SM2} reduces to the action of a D2-brane in an $AdS_4 \times \mathbb{C} P^3$  background. Due to the presence of two dimensionless parameters $N$ and $k$, we can also define a 't~Hooft coupling $\lambda \equiv  N/k$ and consider a 't~Hooft limit $N \rightarrow \infty$ with $\lambda$ fixed. The radius of curvature in string units is of order $\lambda^{1/4}$, so the supergravity description is valid if $\lambda \gg 1$. In this 't~Hooft limit, M-theory (or eleven-dimensional supergravity) always reduces to weakly coupled type IIA string theory (or supergravity), and the spherical M2-branes are really D2-branes. 

From \eq{Veffpert4k}, we can infer that the operator ${\cal O}$ in \eq{O4} scales like $c\sim (kN^3)^{-1/3}$. Since $N/k$ is fixed as $N\to\infty$, the $1/kN^3$ dependence of the triple trace deformation of ABJM theory precisely agrees with the $1/N^4$ scaling assumed in \cite{Craps:2009qc}, based on the requirement that the 't~Hooft limit should exist and be non-trivial.

\setcounter{equation}{0}
\section*{Acknowledgments}
We are grateful to T.~Hertog for collaboration in the early stages of this project, and to V.~Balasubramanian, M.~Berkooz, V.~Hubeny, S.~Minwalla, M.~Rangamani and N.~Turok for useful discussions. We also thank T.~Hertog and N.~Turok for comments on the manuscript. B.C.\ acknowledges the hospitality of the TIFR Monsoon Workshop on String Theory and of the Galileo Galilei Institute for Theoretical Physics as well as partial support from INFN. This work was supported in part by the Belgian Federal Science Policy Office through the Interuniversity Attraction Pole IAP VI/11, by the European Commission FP6 RTN programme MRTN-CT-2004-005104 and by FWO-Vlaanderen through project G.0428.06. 

\appendix

\setcounter{equation}{0}
\section{Brane effective potentials in the Poincar\'e patch}

The computations of D-brane effective potentials in the main text were done for spherical D-branes in global $AdS$ space-times. In this appendix, we discuss the analogous computations for flat D-branes in the Poincar\'e patch of $AdS$. 

In Poincar\'e coordinates, the $AdS_{d+1}$ metric reads
\be
ds^2 = \frac{R^2_{AdS}}{\rho^2} \(-dt^2 + d\rho^2 + d \vec{x}^2\)\,,
\ee
where $d \vec{x}^2$ is the flat metric on $\mathbb{R}^{d-1}$ and $0\le \rho \le \infty$. In this parameterization, the spacetime has an horizon at $\rho = \infty$ and the conformal boundary at $\rho=0$ is $\mathbb{R}^{d-1}$.

Expanding a free massive scalar field in Minkowski plane waves,
\be
\varphi(\vec{x},\rho ) = e^{-i \omega t + i \vec{k}\cdot\vec{x}} \rho^{d/2} \Psi(\rho)\,,
\ee
the radial wave equation becomes
\be
\rho^2 \partial_{\rho}^2 \Psi +\rho \partial_{\rho} \Psi - \left[ m^2 + \frac{d^2}{4} + \rho^2 \( \vec{k^2} - \omega^2 \)\right] \Psi = 0\,.
\ee
For $q^2 = \vec{k}^2 - \omega^2 >0$, the two solutions are \cite{Balasubramanian:1998sn}
\be
\Psi^+_1 (\rho) = K_{\nu}(q \rho)\,, \qquad \Psi^+_2(\rho)= I_{\nu}(q \rho)\,, \label{Poincaresol}
\ee
with $\nu = \frac 1 2\sqrt{d^2 + 4 m^2 R^2_{AdS}}$. In the mass range  $m^2_{BF} \le m^2 < m^2_{BF} + 1/R_{AdS}^2$ we are interested in, corresponding to $0\le\nu <1$, both solutions are normalizable at the boundary of spacetime, while only $\Psi^+_1$ is regular in the interior.%
\footnote{
For $q^2 < 0$, the solutions are 
\be
\Psi^-_{1/2}  (\rho) = J_{\pm \nu}(|q| \rho)\,,
\ee
when $\nu$ is non integer and
\be
\Psi^-_1 (\rho) = J_{\nu}(|q| \rho)\,, \qquad \Psi^-_2(\rho)= Y_{\nu}(|q| \rho)
\ee
for integer $\nu$.
} 
In our computation of the D-brane effective potential, one would expect that only the $q^2=0$ modes contribute. However, we will see that it is useful to consider a regulator momentum $q_0^2>0$. For $q^2>0$, we construct the propagator starting from the solution that is regular at the origin and from a solution with specified behavior near the boundary: 
\ba
\Psi_1 (\rho) =  K_{\nu}(q \rho)\,, \qquad \Psi_2(\rho) =  I_{\nu}(q \rho) + C^P_{\infty} K_{\nu}(q \rho)\,,
\ea
where $C^P_{\infty}$ will be chosen such that $\Psi_2$ satisfies the boundary conditions of interest.
The Feynman propagator then reads
\ba
G_F (\vec{x},\rho;\vec{x}^{\prime},\rho^{\prime}) &= & - \frac{\kappa_{D}^2}{R^{(d-1)}_{AdS} V_{S^{D-(d+1)}}} \int_{-\infty}^{\infty} \frac{d\omega}{2\pi} \int \frac{d^d \vec{k}}{(2\pi)^d} e^{-i \omega(t-t^{\prime})+ i \vec{k} \cdot (\vec{x} - \vec{x}^{\prime})}\times  \nonumber\\ 
&&\rho^{d/2}  \rho^{\prime d/ 2} \{ \theta(\rho -\rho^{\prime})  \Psi_1 (\rho) \Psi_2 (\rho^{\prime})+ \theta(\rho^{\prime}-\rho) (\rho \leftrightarrow \rho^{\prime})\}\,,
\ea
where $D=10$ (or $D=11$) respectively for type IIB supergravity (or eleven-dimensional supergravity).
We are now ready to specialize to the two cases of interest in this paper. Consider a probe D3-brane (or M2-brane) extended in flat four-dimensional (or three-dimensional) space sitting at a radial Poincar\'e coordinate $\bar{\rho}$ and localized at a point in $S^5$ (or $S^7$).

In the five-dimensional setup of section~2, $\nu =0$ and the source terms are
\be
\mathcal{J}_{DBI}(\rho) = 5\gamma \frac{\tau_3}{R_{AdS}}  \(\cos^2 \xi - \frac 1 5 \sin^2 \xi\)\rho \,\delta(\rho-\bar{\rho} )\,, \label{JDBI5P}
\ee
\ba
\mathcal{J}_{WZ} (\rho) = \left\{ \begin{array}{ll} \label{JWZ5P}
- 10 \gamma \frac{\mu_3 }{R_{AdS}}  (\cos^2 \xi - \frac 1 5 \sin^2 \xi)  & \textrm{$\rho \ge \bar{\rho} $}\\ 
0 & \textrm{$\rho  < \bar{\rho} $}\,.
\end{array} \right. 
\ea
The propagator satisfying the boundary conditions \eq{BCs} defined at the scale $\mu$ appearing in \eq{scalar}, has
\be
C^P_{\infty} =\frac{f}{1 + f\( \gamma_E +  \ln \frac{q\mu R_{AdS}^2}{2}\)}\,,
\ee
where $\gamma_E$ is again Euler's constant. Here we see why it is useful to introduce a regulator $q^2=q_0^2>0$: for $q^2=0$, we would have found an infrared divergent expression (we will comment more on this below).
The (regularized) effective potential computed as in (\ref{Veffprop}) is
\be\label{VeffP}
\int d^4x \,V_{eff} ( x) = - \frac{f}{1 + f\( \gamma_E +  \ln \frac{q_0\mu R_{AdS}^2}{2}\)}  \frac{5 \pi^2}{ 3 N^2} \int d^4 x\, \left[ \phi_1^2 - \frac 1 5 \sum_{i=2}^6 \phi_i^2 \right]^2\,,
\ee
where we have introduced the fields 
\be \label{PhiPoincare}
\phi_1 \equiv \sqrt{\tau_3} \frac{ R^2_{AdS}}{\bar \rho}  \cos \xi, \ \ \ \phi_2 \equiv  \sqrt{\tau_3}  \frac{ R^2_{AdS}}{\bar \rho}  \, \sin\xi \cos \Omega_1,\ \ \ \ldots
\ee
with canonical kinetic term
\be
S_{kin} = - \frac 1 2 \int d^4 x\, \partial_{\alpha} \phi_i \partial^{\alpha} \phi_i\,.
\ee
We can now explain what is the role of the IR regulator $q_0^2$. Since the sources do not depend on $t$ and $\vec x$, only the $q^2=0$ modes should contribute to the effective potential. As is easy to see by letting $q_0 \rightarrow 0$ in (\ref{VeffP}), this would formally give a vanishing result. From a dual field theory point of view, this can be understood as follows. The scale $1/(\mu R_{AdS}^2)$ corresponds to the scale at which the coupling constant $f$ is defined, while $q$ is the scale at which the (renormalized) four-point function is computed. In the planar (large $N$) limit, factorizable diagrams such figure~\ref{figHigher} can be resummed and give rise to the running coupling $f/[1 + f(\gamma_E +  \ln(q_0\mu R_{AdS}^2/2))]$ appearing in \eq{VeffP}. Note in particular that the formal vanishing of the coupling for $q_0\to 0$ is not reliable: for $f>0$, the coupling becomes strong as one flows to the IR and formally becomes infinite at some finite value of $q_0$, before $q_0=0$ is reached. Note also that these infrared divergences were absent in section~2.4, since there the radial position of the brane was effectively massive (corresponding to the conformal coupling to the curvature of $S^3$ in SYM theory on $\mathbb{R}\times S^3$).

In the four-dimensional case, $\nu = 1/2$ and the sources read
\be
\mathcal{J}_{DBI}(\rho) = - \frac{\tau_2 }{R_{AdS}} \frac{1}{\sqrt 2} \(\cos^2 \theta -  \sin^2 \theta\)  \rho \, \delta(\rho-\bar{\rho})\,, \label{JDBI4P}
\ee
\ba
\mathcal{J}_{WZ} (\rho) = \left\{ \begin{array}{ll} \label{JWZ4P}
2 \frac{ \mu_2 }{R_{AdS}} \frac{1}{\sqrt 2} (\cos^2 \theta - \sin^2 \theta)  & \textrm{$\rho \ge \bar{\rho} $}\\ 
0 & \textrm{$\rho  < \bar{\rho} $}\,.
\end{array} \right. 
\ea
The supersymmetric boundary condition sets 
\be
C^P_{\infty} = \frac{2}{\pi}\,.
\ee
The result for the effective potential computed as in (\ref{Veffpert4}) is
\be
\int d^4 x\, V_{eff}(x) = -\frac {h}{N^3} \frac{3 \pi^2}{8} \int d^4 x\, \left[ \sum_{i=1}^4 \phi_i^2 - \sum_{i=5}^8 \phi_i^2 \right]^3\,,
\ee
in terms of the canonically normalized scalars
\begin{eqnarray}
\phi_1 &\equiv& 2 \sqrt{\frac{\tau_2 R^3_{AdS}}{ \bar \rho}} \cos \theta  \cos \Omega_1, \  \ \ \phi_2 \equiv 2 \sqrt{\frac{\tau_2 R^3_{AdS}}{ \bar \rho}} \cos \theta  \sin \Omega_1 \cos \Omega_2, \ \ \dots \nonumber \\ 
\phi_5 &\equiv& 2 \sqrt{\frac{\tau_2 R^3_{AdS}}{ \bar \rho}} \sin \theta  \cos \Omega_4, \ \ \  \phi_6 \equiv 2 \sqrt{\frac{\tau_2 R^3_{AdS}}{ \bar \rho}} \sin \theta  \sin \Omega_4 \cos \Omega_5, \ \ \dots
\end{eqnarray}
The final result does not depend on the regulator $q_0$. This is in agreement with the fact that the boundary conditions \eq{BCs4} are $AdS$ invariant and that in the planar limit the corresponding multi-trace deformation is exactly marginal and preserves conformal invariance. 


\end{document}